\documentclass[aps,prl,twocolumn,floats,showpacs]{revtex4}

\usepackage{graphicx}
\usepackage{amsmath}
\usepackage{amssymb}
\usepackage[usenames,dvipsnames]{color}

\newcommand{\w}{{\omega}}

\renewcommand{\l}{{\lambda}}

\newcommand{\X} {{X^1\Sigma_{0 g}^+}}
\newcommand{\B} {{B^3\Pi_{0+u}}}

\renewcommand{\wp}{{\omega_p}}
\newcommand{\ws}{{\omega_S}}

\newcommand{\Sch}{{Schr\"{o}dinger }}

\newcommand{\beq}{\begin{equation}}
\newcommand{\eeq}{\end{equation}}
\newcommand{\bea}{\begin{eqnarray}}
\newcommand{\eea}{\end{eqnarray}}

\begin{document}
\title{Enhancing strong-field induced molecular vibration with femtosecond pulse shaping}

\date{\today}
\author{M.~Bitter$^{1}$, E.~A.~Shapiro$^{2}$  and V.~ Milner$^{1}$}
\affiliation{Department of  Physics \& Astronomy and The Laboratory for Advanced Spectroscopy and Imaging Research (LASIR)$^{1}$, and Department of Chemistry$^{2}$, The University of British Columbia, Vancouver, Canada \\}

\begin{abstract}{
This work investigates the utility of femtosecond pulse shaping in increasing the efficiency of Raman excitation of molecules in the strong-field interaction regime. We study experimentally and theoretically the effect of pulse shaping on the strength of non-resonant coherent anti-Stokes Raman scattering in iodine vapor at laser intensities exceeding $10^{13}$ W/cm$^2$. We show that unlike the perturbative case, shaping strong non-resonant laser pulses can increase the signal strength beyond that observed with the transform-limited excitation. Both adiabatic and non-adiabatic schemes of excitation are explored, and the differences of their potential in increasing the excitation efficiency are discussed.
}
\end{abstract}

\maketitle


\section{Introduction}
The broad spectral bandwidth of ultrashort laser pulses is often
used to excite coherent molecular wave packets consisting of a
number of rotational and vibrational eigenstates. Studying the
dynamics of an excited wave packet on femtosecond time scale
represents a popular approach to molecular spectroscopy
\cite{Zewail-book}. Femtosecond spectroscopy benefits from the
technology of pulse shaping \cite{Weiner2000} which offers
selectivity and control of molecular excitation and has been
successfully implemented in various applications such as
multiphoton ionization \cite{Pearson2001}, stimulated Raman
scattering \cite{Weiner1990,Pearson2001}, four-wave mixing
\cite{Hornung2001,Hauer2006b}, coherent anti-Stokes Raman
scattering
\cite{Scully2002,Silberberg2009,Konorov2009,Zheltikov2000,Hellerer2004,Zeidler2002,Li2010}
and coherent control of chemical reactions \cite{Lozovoy2006}.

In the weak-field regime of Raman excitation of molecules, i.e. when molecular states are not changed significantly by the applied laser fields and the perturbation theory holds, the complex amplitudes of the excited molecular states are proportional to the corresponding resonant spectral components of the two-photon excitation field \cite{Meshulach1998,Meshulach1999}. In the absence of intermediate resonances, an upper bound on the \textit{absolute efficiency} of exciting a particular mode is set by the available laser intensity and is reached with unshaped transform-limited (TL) laser pulses. Hence, improving the efficiency of an off-resonance Raman process in the perturbative regime can be achieved by increasing the laser intensity, whereas shaping the spectrum of the driving field can only suppress rather than enhance the excitation at any given frequency. Stronger laser fields modify the molecular field-free states, most importantly via AC Stark shifts, often suppressing the rate of the target transition \cite{Trallero-Herrero2006, Trallero-Herrero2007, Konorov2011, Zhdanovich2011}. As a result, unshaped pulses no longer provide maximum efficiency of transfering molecules to the target vibrational state in the strong-field limit.

In this work, we investigate the utility of femtosecond pulse shaping to increase the magnitude of non-resonant vibrational excitation.  We explore, both experimentally and theoretically, the efficiency of exciting vibrational wave packets in the ground electronic state of molecular iodine subject to strong laser pulses ($>10^{13}$ W/cm$^2$). Evaluating the efficiency of strong-field induced vibrational excitation by means of coherent anti-Stokes Raman scattering (CARS), we demonstrate that pulse shaping can lead to the enhancement of the nonlinear spectroscopic signal by at least 50\%.

Two qualitatively different approaches are analyzed. First, we suppress strong-field effects (such as AC Stark shifts) that may reduce the efficiency of the CARS process. In contrast to the feedback-loop based adaptive control \cite{Judson1992,Scully2002,Trallero-Herrero2006}, we achieve this by applying pre-determined pulse shapes to the excitation pulses in such a way as to lower the instantaneous field strength while preserving the amplitude of the two-photon field at the frequency of Raman resonances.

In the second approach we try to employ the technique of adiabatic passage (AP) to enhance off-resonance vibrational excitation by making use of, rather than avoiding, strong-field effects. Non-adiabatic population transfer is very sensitive to the fluctuations of the laser parameters like intensity or pulse duration \cite{Vitanov2001}. In AP, one exploits an adiabatic time evolution to improve the robustness of the process and enhance the efficiency of population transfer. Earlier works have demonstrated a number of successful implementations of adiabatic transfer in two- and three-level systems driven by resonant laser radiation \cite{Gaubatz1990,Broers1992,Melinger1994,Sautenkov2004,Zhdanovich2008}. Previous studies questioned the utility of AP in multi-level molecular systems interacting with strong femtosecond pulses \cite{Graefe2007,Graefe2007b}. In this work we put to test the techniques of two-photon rapid adiabatic passage (RAP \cite{Oreg1984, Malinovsky2001}), stimulated rapid adiabatic passage (STIRAP \cite{Gaubatz1990}) and ``piecewise AP'' \cite{Shapiro2007a,Shapiro2008}. The latter approach has been proven successful in atomic systems \cite{Zhdanovich2008, Zhdanovich2009} but has not been implemented with molecules.

We use femtosecond CARS to determine the efficiency of vibrational excitation in the ground electronic state of molecular iodine. Two strong laser pulses, pump and Stokes, with frequencies $\omega_\text{p}$ and $\omega_\text{S}$ and intensities exceeding $10^{13}$ W/cm$^2$, prepare the vibrational coherence via an off-resonance two-photon Raman transition. A weak probe pulse with frequency $\omega_{pr}$, separated in time from pump and Stokes, scatters off this coherence generating the anti-Stokes field at $\omega_{aS}=\omega_{p}-\omega_{S}+\omega_{pr}$. The detected CARS signal at $\omega_{aS}$ serves as a quantitative measure of the degree of vibrational excitations, and its dependence on the spectral shapes of both pump and Stokes pulses is explored.

\section{Choice of pulse shaping}
The spectral phase of both pump and Stokes excitation fields is
shaped simultaneously in such a way as to (1) suppress the
strong-field induced level shifts, or (2) initiate an AP process.
In both cases, we aim at increasing the strength of the CARS
signal. The important difference between the two approaches lies
in the two-photon  pump-Stokes field. In scenario one, the
spectral power densities at the desired Raman transition
frequencies are maintained while the peak intensities of the
pulses are reduced. Thus, the detrimental strong-field Stark
shifts are minimized. In the second scenario, not only the field
amplitudes of undesired frequencies but also those of the target
Raman transitions are lower than in the TL case.
\begin{figure}
\centering
 \includegraphics[width=1\columnwidth]{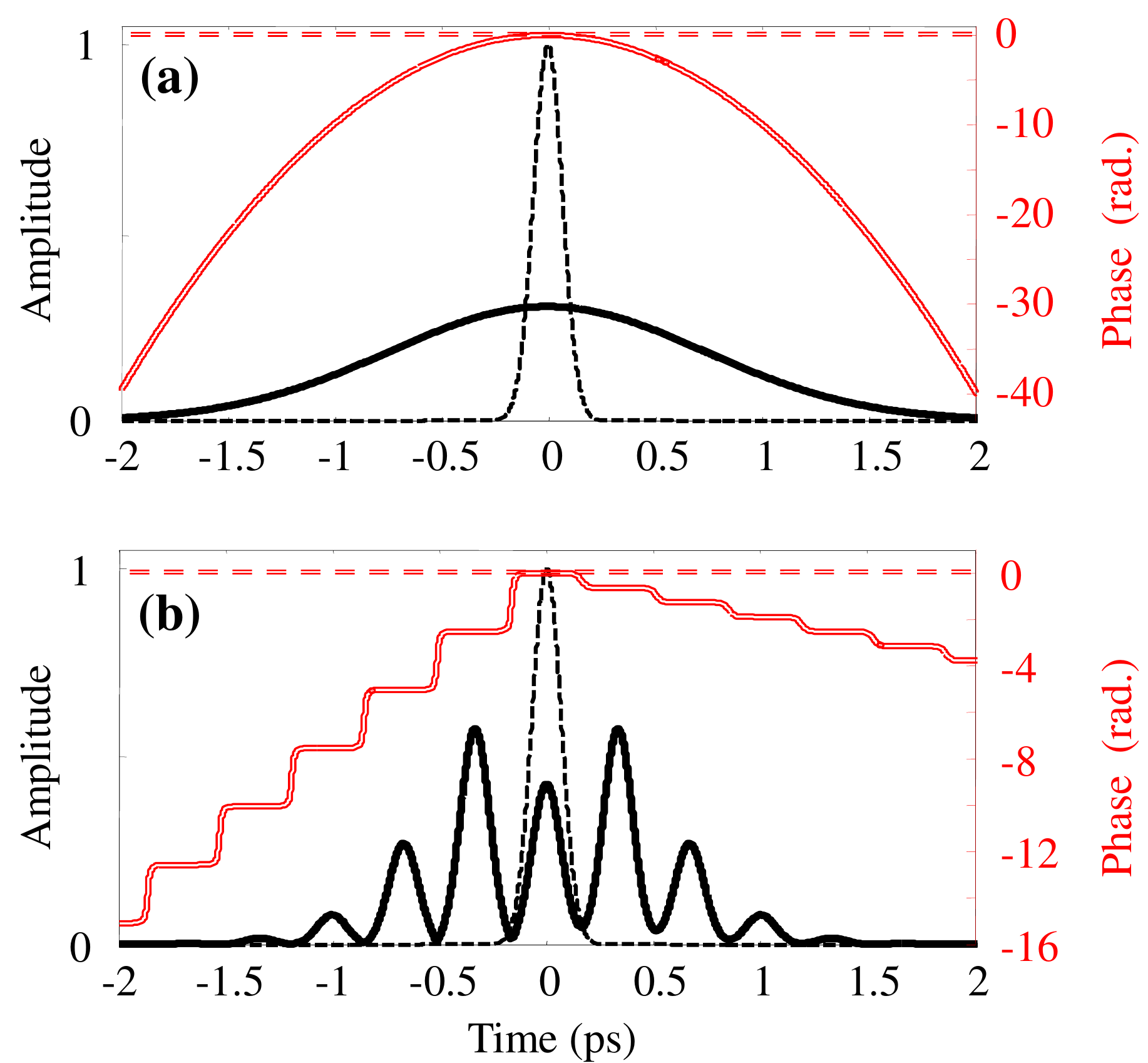}
     \caption{(Color online) Temporal profile of TL (dashed lines) and shaped (solid lines) femtosecond pulses with the electric field amplitude (black single lines) and phase (red double lines). (a) Linearly chirped pulse with $\alpha'=50,000$ fs$^2$, (b) sinusoidal phase modulation with $A=1.65$, $T=334$ fs.}
  \vskip -.1truein
  \label{Fig:PulsesTimeDomain}
\end{figure}

\subsection{Linear frequency chirping}
An equal frequency chirp applied to both pump and Stokes pulses results in the narrowing of the two-photon excitation spectrum around a constant frequency which is tuned to the frequency of the target vibrational transition $\Omega_0$. The spectral power density at that Raman transition is kept constant, whereas the peak power of the individual pulses can be substantially lowered.

\begin{figure}
\centering
 \includegraphics[width=0.9\columnwidth]{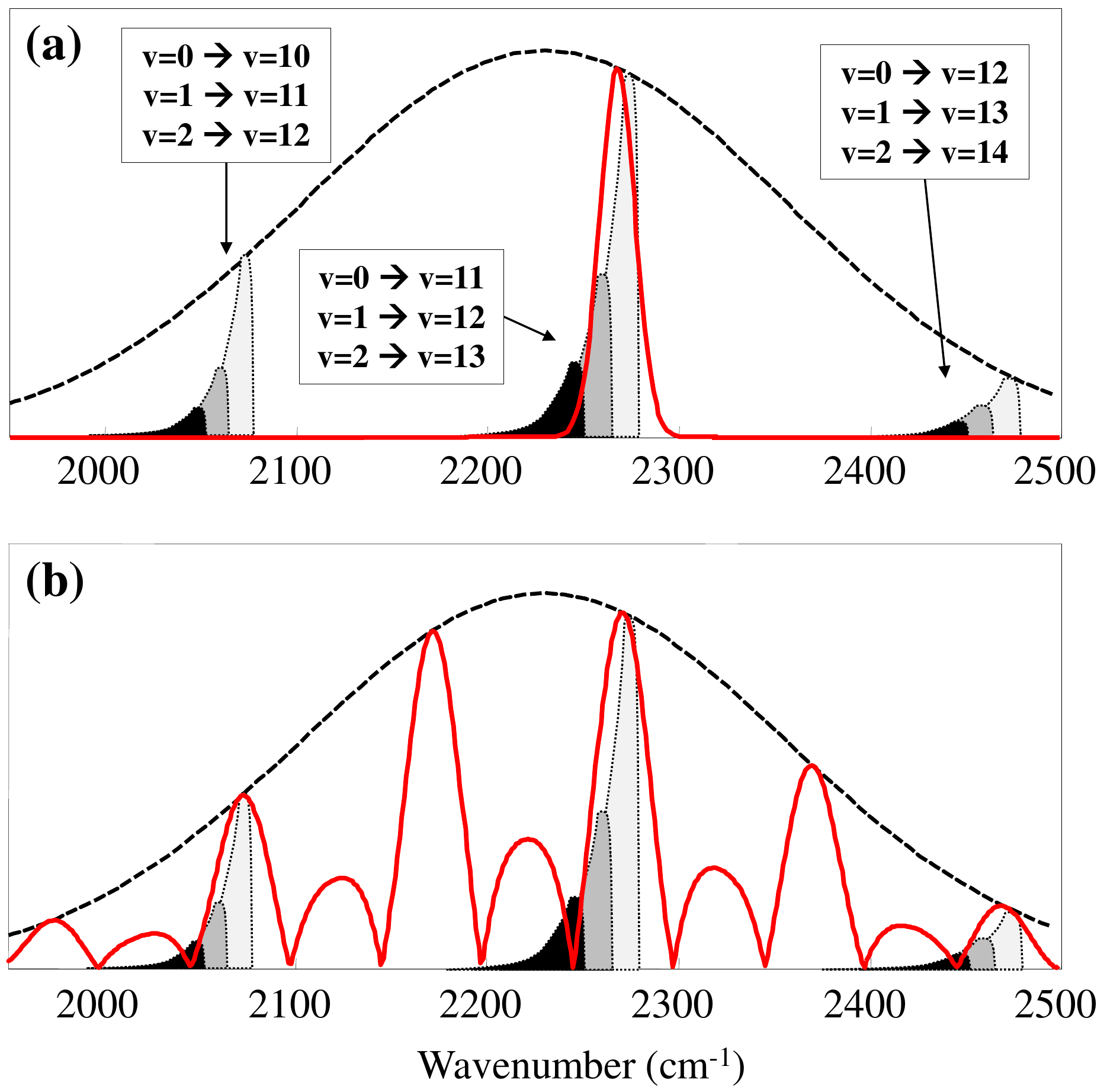}
  \caption{(Color online) Raman spectrum at $\wp-\ws$  for TL pulses (black dashed line) and shaped pulses (red solid line). All Raman transitions (rotationally broadened into bands) and their strengths are indicated according to thermal Boltzmann distribution at 100 $^\circ$C: initial states v=0 (white area), v=1 (grey area) and v=2 (black area). (a) Linearly chirped pulses with $\alpha'=50,000$~fs$^2$, (b) sinusoidal phase modulation with $A=1.65$, $T=334$~fs. } \vskip -.1truein
  \label{Fig:RamanSpectrum}
\end{figure}

Linear chirping of the instantaneous frequency of pump and Stokes fields $\w_{p,S}(t)=\w_{0(p,S)} +\alpha t$ is achieved by applying the parabolic spectral phase masks by means of two pulse shapers: \begin{equation}
    \varphi_{p,S}(\w)= - \frac{1}{2}\alpha' (\w-\w_{0(p,S)})^2 .
\end{equation}
The magnitude of the spectral chirp $\alpha'$ is related to the temporal chirp $\alpha = 1/\alpha'$ \cite{Malinovsky2001}. $\omega _{0p}$ and $\omega _{0S}$ are the central frequencies of pump and Stokes pulses, respectively. An example of the temporal amplitude and phase before and after linear chirping is shown in Fig.~\ref{Fig:PulsesTimeDomain}(a). In comparison to a transform-limited pulse, the peak amplitude of the shaped pulse drops as the pulse is stretched in time in linear proportion with $\alpha'$.
The instantaneous frequency of the two-photon excitation spectrum of two linearly chirped pump and Stokes pulses with an equal chirp rate $\alpha$ remains constant, $\Omega (t) := \omega _{p}(t) - \omega _{S}(t) = \omega _{0p} - \omega_{0S}$, as shown in Fig.~\ref{Fig:RamanSpectrum}(a). Together with the two-photon field spectrum, we plot all possible Raman transitions in room-temperature iodine vapor within the accessible spectral bandwidth. Initial thermal population distribution among the vibrational levels $v=0,1,2,...$ is taken into account. Vibrational resonances are broadened due to the thermal molecular rotation. We do not resolve the rotational spectrum in our experiments. The excitation line width is inversely proportional to the applied chirp $\alpha'$, whereas its frequency can be tuned by a variable time delay between pump and Stokes pulses. In CARS microscopy this method is known as spectral focusing \cite{Hellerer2004}. In our experiments, $\alpha '=50,000$ fs$^2$ is chosen to match the excitation line width with the rotational broadening of the vibrational transitions.

\subsection{Sinusoidal phase modulation}
Narrowing of the two-photon spectrum around the target Raman frequency $\Omega _{0}$ requires the spectral phase of pump and Stokes pulses to be identical around $\omega _{0p}$ and $\omega _{0S}=\omega _{0p}-\Omega_{0}$, respectively, as in the case of an equal frequency chirp described above. If the goal is to excite a vibrational wave packet consisting of several eigenstates, the applied phase mask must be periodic in frequency, with a period matching the vibrational period of a molecule \cite{Weiner1990,Hauer2006a, Hauer2006b}.

Sinusoidal phase modulation represents one of the most popular methods of periodic shaping \cite{Wollenhaupt2006, Shapiro2007a, Shapiro2008,Voll2009}. In this case,
\begin{equation}
    \varphi_{p,S}(\omega)=A\sin[~(\w-\tilde{\w}_{p,S})T~],
    \label{SinusoidalPhase}
\end{equation}
where $T$ describes the modulation period and $A$ its amplitude.
$\tilde{\w}_{p}$ and $\tilde{\w}_{S}$ are the central modulation
frequencies of pump and Stokes pulses, respectively. In the time
domain, this spectral shaping results in a train of pulses, with
each pulse being an exact replica of the initial transform-limited
pulse, and the time delay between the pulses defined by $T$
(Fig.~\ref{Fig:PulsesTimeDomain}(b)). By matching or mismatching
the train period with the period of molecular vibration, the
latter can be either enhanced or suppressed. Weak-field
vibrational control by means of periodic pulse trains has been
successfully demonstrated \cite{Hauer2006a, Hauer2006b,
Konradi2006,Konorov2009}. In the strong-field regime considered in
this work, breaking the initial pulse into a train of weaker
pulses serves the purpose of suppressing the detrimental
multi-photon processes prohibiting the transfer of molecules to
the target state.

If an identical sinusoidal phase modulation is applied to both pump and Stokes excitation pulses, the two-photon spectrum shows periodic peaks with flat phase across them, see Fig.~\ref{Fig:RamanSpectrum}(b). Unlike the case of a frequency chirp, all three vibrational bands are now excited with the maximum possible amplitude (set by the available bandwidth) if the period $T$ is a multiple integer of the vibrational period. The modulation amplitude $A$ determines the number of pulses in the pulse train, as well as the line width of the spectral peaks in the two-photon spectrum \cite{Wollenhaupt2006}. Similarly to the parabolic phase shaping described above, we tune the value of $A$ for the best coverage of the rotationally broadened Raman transitions in iodine. The central modulation frequencies $\tilde{w}_{p,S}$ in Eq.~\ref{SinusoidalPhase} control the absolute position of the peaks in the two-photon spectrum.
\subsection{Pulse shaping for Adiabatic Passage}
In RAP, the instantaneous frequency of the excitation field is swept across the target transition frequency \cite{Malinovsky2001,book-Abragam}. In the case of a Raman process, this corresponds  to  chirping the two-photon frequency. Applying linear frequency chirps with opposite signs to pump and Stokes fields $\alpha_p=-\alpha_S=\alpha$, results in a linearly chirped two-photon field with an instantaneous frequency of $\Omega (t) = \omega _{p}(t) - \omega _{S}(t) = \omega _{0p} - \omega_{0S} +2\alpha t$.

The technique of STIRAP relies solely on a relative time delay between pump and Stokes pulses.  Hence, no pulse shaping is necessary. The pulse trains needed for Piecewise STIRAP can be implemented via amplitude and phase shaping according to the method described in Ref.~\cite{Zhdanovich2008} (see also Fig.~\ref{Fig-IntensityScan-AP}).

\section{Numerical analysis}
In our experimental detection scheme based on CARS, the signal
from a \textit{single} final state $|v_f\rangle$ is proportional
to the square of the coherence between that state and the initial
vibrational state $|v_0\rangle$, i.e. the square amplitude of
$|v_0\rangle\langle v_f|$ in the density matrix. Since the rate of
both collisional and rotational decoherence is negligible on the
experimentally realized time scale of a few picoseconds, and the
target population is small, the signal magnitude is proportional
to the population of the final state. We therefore base our
numerical analysis on calculating the transfer of population from
the ground vibrational state, predominantly populated at our
experimental conditions, to higher vibrational states, driven by
strong pump and Stokes excitation fields. For a {\it coherent wave
packet} consisting of several vibrational final states, the
observed CARS signal oscillates as a function of the arrival time
of probe pulses (see Results section below). Rather than model
this time behavior, we assume that the time-averaged signal is
proportional to the sum of the calculated populations.
To calculate the re-distribution of the vibrational population as
a result of the strong-field interaction, we solve the
time-dependent \Sch equation in the eigenstate basis. The
propagation routine does not employ the rotating wave
approximmation and thus automatically accounts for AC Stark
shifts. We take into account transitions between two electronic
manifolds of $I_{2}$, $\X$ and $\B$. Both potentials are modeled
as Morse oscillators \cite{NIST}. The electronic transition dipole
moment is set to an approximate value of 1 Debye, while the
Franck-Condon factors for the $B \leftarrow X$ transitions are
calculated numerically. While such modelling is not fully accurate
for high vibrational states of iodine \cite{Zare1964}, it can
provide a good qualitative description of the quantum dynamics.
Quantitatively, in our simulations the strong-field effects arise
at intensities somewhat lower than those observed experimentally.

In the main set of calculations, we explore the simplest case of a strongly driven off-resonance Raman excitation. We avoid one-photon resonances, and set the central wavelength of the pump pulse to $\l_p = 800$ nm whereas the Stokes pulse is varied between $\l_S$ = 940 nm and 990 nm. Importantly, neither pump nor Stokes photons have enough energy to drive a resonant transition between either $|v_0\rangle$ or $|v_f\rangle$ and the vibrational states of the $B$ manifold. Various multi-photon interaction pathways that couple $|v_0\rangle$ to other electronic and vibrational states become intertwined in the strong-field limit. The population transfer is influenced by the time-dependent Stark shifts which move various vibrational levels in and out of resonance with different components of the Raman excitation spectrum. Nonlinear strong-field processes, most notably dynamic Stark shifts, make the perturbative approach inapplicable and lead to the observed suppression of CARS signal.

We first focus on the pulse shapes designed to reduce the peak intensity of the excitation pulses, while enabling the excitation of either a single vibrational level (quadratic spectral phase mask) or a superposition of levels (sinusoidal spectral phase mask). In the former case, a linear chirp with $\alpha' = 46,000$ fs$^2$ stretches pump and Stokes pulses to $\sim 1$ ps duration. In the second case, the sinusoidal mask of Eq.(\ref{SinusoidalPhase}) with $A=1.23$ and $T=312$ fs leads to a pulse train consisting of five pulses separated in time by twice the oscillation period of a wave packet composed of the vibrational states adjacent to $v_f = 10$. The factor of 2 is introduced for an easier distinction of individual pulses in the resulting pulse train.

\begin{figure}
\centering
 \includegraphics[width=0.8\columnwidth]{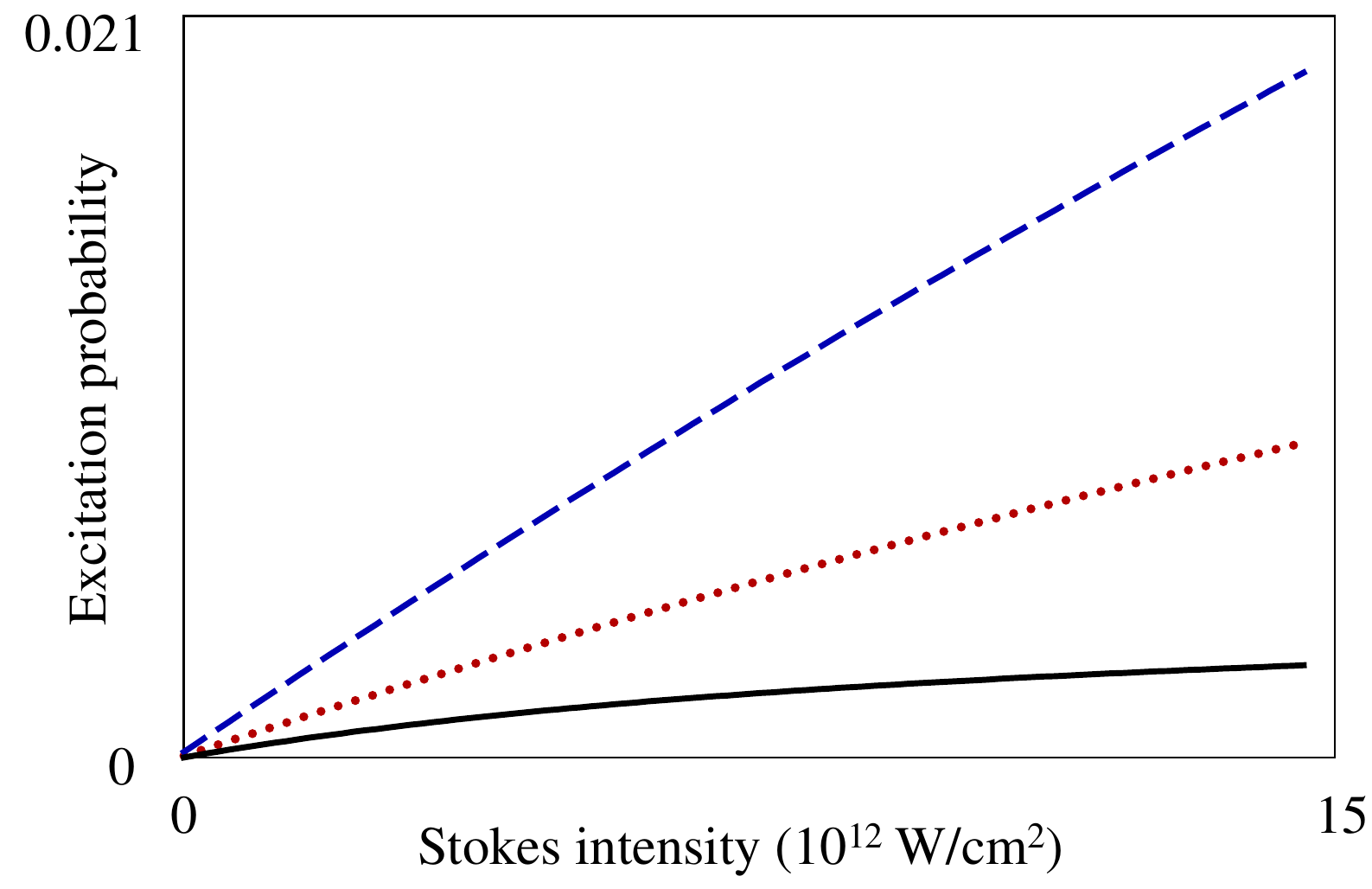}
     \caption{(Color online) Raman excitation probability as a function of Stokes intensity. The excitation schemes are: TL pulses (black solid line), pump and Stokes pulses are linearly chirped (blue dashed line), pump and Stokes pulses are shaped with a sinusoidal phase mask (red dotted line).
  }
  \vskip -.1truein
  \label{FigZ-IntensityScan}
\end{figure}

Figure \ref{FigZ-IntensityScan} demonstrates the dependence of the
Raman transfer efficiency on laser intensity. Here, pump intensity
($I_{p}$) is fixed at $1.5\times 10^{13}$ W/cm$^2$  while Stokes
intensity ($I_{S}$) is scanned. The Stokes wavelength is set to
964 nm corresponding to a strong two-photon resonance with
$v_f=10$. We compare the effect of 130-fs $\sin^2 t/t_0$-shaped TL
pump and Stokes pulses (full width at half maximum (FWHM) of the
intensity profile) and that of the shaped pulses simulated with a
numerical pulse shaper. As the experimental detection is carried
out with a broadband probe and therefore does not offer frequency
resolution, we plot the sum of populations in the target manifold,
$v=6..13$, rather than the populations of the individual levels.
This range is wide enough to include all the states covered by the
excitation bandwidth in the presence of strong-field Stark shifts.
On the other hand, it does not include those states which can be
populated in pump-pump and Stokes-Stokes Raman transitions in the
presence of Stark shifts.

The lower curve in Fig.~\ref{FigZ-IntensityScan} illustrates the
main motivation of this work: the efficiency of the Raman
excitation driven by high-intensity transform-limited pulses
saturates, setting an upper bound on the magnitude of the CARS
signal. The saturation stems from an oscillatory behavior of the
individual target populations, which takes over the linear (with
$I_{S}$) growth of population anticipated in the weak-field limit.
We associate these population oscillations with an interplay of
population transfer with multiple AC Stark shifts which
dynamically change the instantaneous spectrum of the molecular
states dressed by the strong laser field. We note that on the
scale of laser intensities considered in this work, the
oscillation amplitudes are typically of the order of 1-2\% of the
total population, which explains the low total efficiency of the
Raman excitation plotted in Fig.~\ref{FigZ-IntensityScan}. If the
two-photon field is resonant with a Raman transition to a certain
vibrational state $|v_f\rangle$, then the linear frequency
chirping of pump and Stokes pulses provides the largest
suppression of the strong-field saturation (dashed curve in
Fig.~\ref{FigZ-IntensityScan}). Raman excitation by a train of
pump/Stokes pulse pairs (sinusoidal spectral mask) showed an
intermediate level of saturation (dotted curve in
Fig.~\ref{FigZ-IntensityScan}). We attribute this difference in
performance to the fact that for a given pulse energy, frequency
chirping results in the lowest peak intensity and therefore better
suppression of the detrimental strong-field effects discussed
above.

\begin{figure}
\centering
 \includegraphics[width=1\columnwidth]{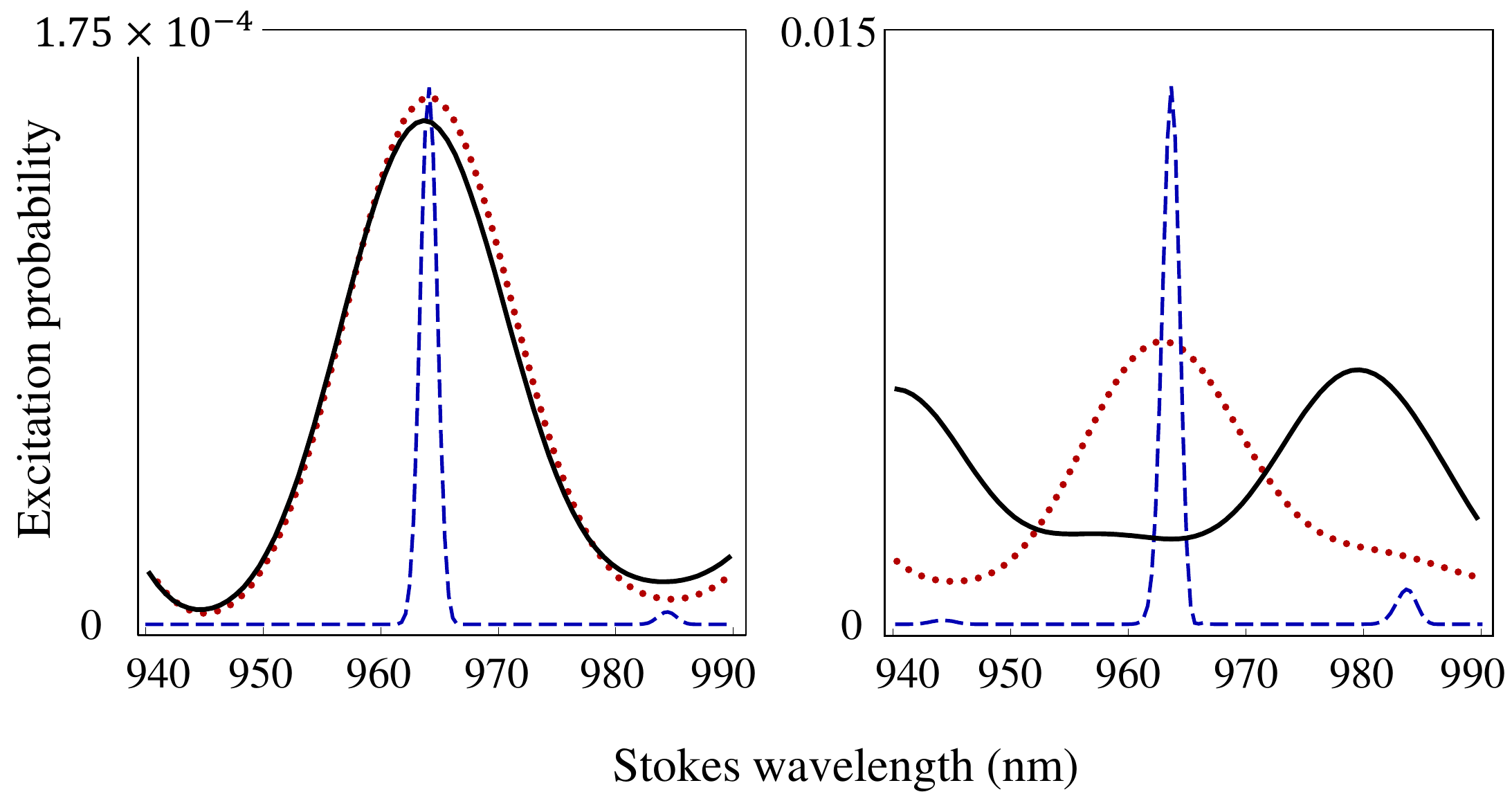}
     \caption{(Color online) Raman excitation probability as a function of Stokes wavelength. Left: both pump and Stokes intensities are set to   $10^{12}$ W/cm$^2$. Right: both intensities are set to  $10^{13}$ W/cm$^2$. Line assignment is the same as in Fig.~\ref{FigZ-IntensityScan}.}
  \vskip -.1truein
  \label{FigZ-FrequencyScan}
\end{figure}

Further insight can be gained by calculating the transfer efficiency as a function of the central Stokes frequency, $\omega _{0S}$. The results are shown in Fig.~\ref{FigZ-FrequencyScan}. In the weak-field regime (left panel), the transition strength is defined by the corresponding resonant component of the two-photon pump-Stokes spectrum. Since the peak spectral amplitudes are equal for all three pulse shapes, the population transfer to a single state $|v_f\rangle$ is equally efficient when $\w_{0p}-\w_{0S} = (E_f-E_0)/\hbar$, where $E_{k}$ is the energy of the $k$-th vibrational state. In this figure, the Raman resonance with $v_{f}=10$ lies at $\l_{S}=964$ nm. Raman transitions to the neighboring $v_f=9$ and $v_f=11$ are much weaker due to oscillations in the two-photon coupling strength. For the frequency chirped pulses, the two-photon spectrum is narrower than in the case of TL pulses (see Fig.~\ref{Fig:RamanSpectrum}(a)), resulting in a faster drop of the transfer efficiency away from the resonance. The two-photon spectrum of the pump-Stokes pulse train has a much wider maximum \footnote{Raman line width for a train of pump-Stokes pulse pairs depends on the relation between $\tilde{\w}_{p,S}$ in Eq.\ref{SinusoidalPhase} and the central pump and Stokes frequencies $\w_{0(p,S)}$. In the calculation shown in Fig.4, the central Stokes wavelength is scanned while $\tilde{\w}_{S}$ is fixed. In this scenario, the Raman line width at small intensities is as wide as that for a transform-limited pulse.}.

The right panel of Fig.~\ref{FigZ-FrequencyScan} shows the calculated transfer efficiency in the strong-field limit. Dynamic Stark shifts modify the line shape and lead to a significantly lower population of the target vibrational manifold. The vibrational excitation by a train of pulses is suppressed to a much lower degree. Indeed, the intensity of pulses in the train is lower than that of the original femtosecond pulse and the strong-field effects are correspondingly weaker. Finally, chirped pulses correspond to the highest population transfer. The absence of a visible line broadening in this case points to the complete elimination of the detrimental strong-field effects.

\begin{figure}
\centering
\includegraphics[width=1\columnwidth]{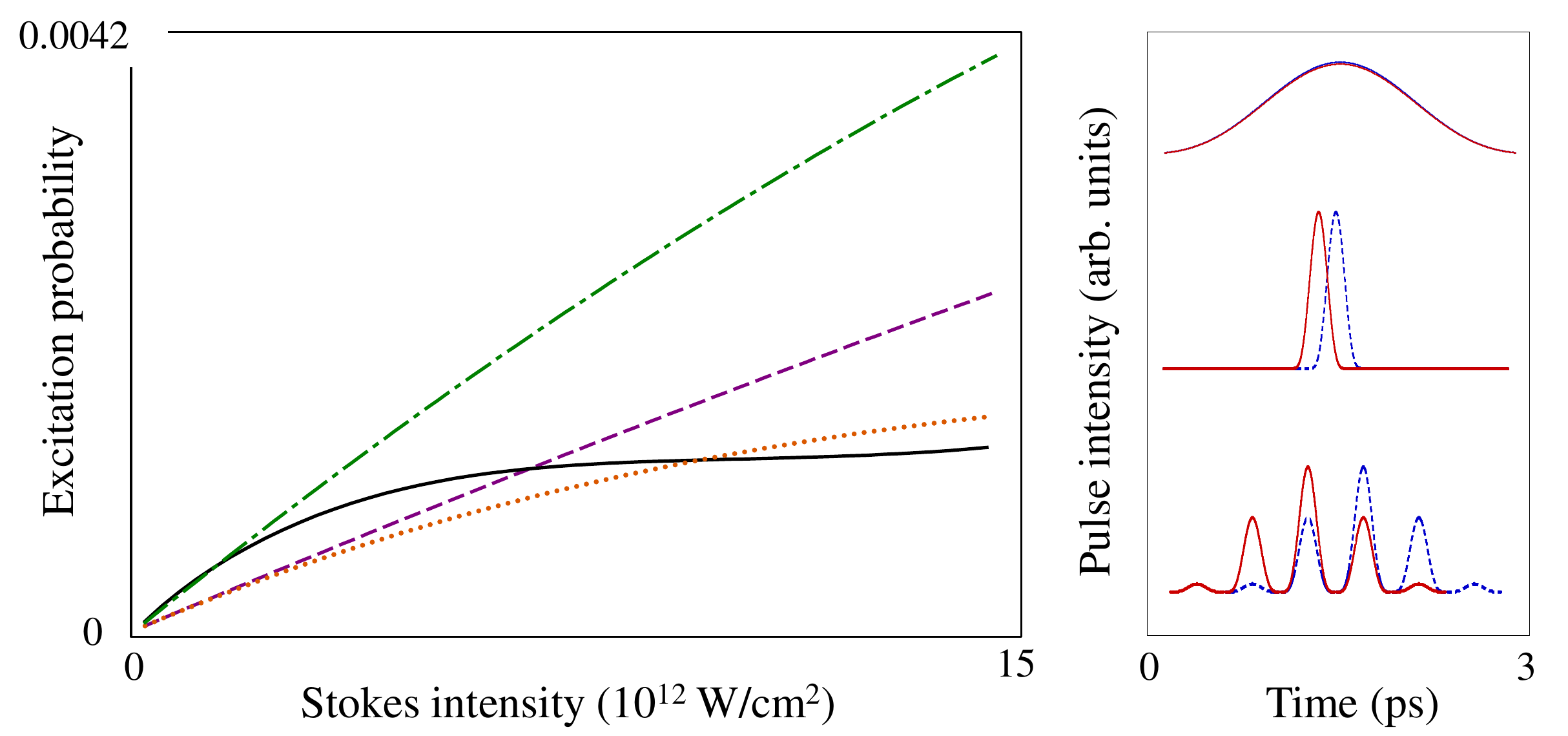} \caption{(Color online) Left: Raman excitation probability as a function of Stokes intensity. The excitation schemes are: RAP (purple dashed line), STIRAP (orange dotted line), piecewise STIRAP (green dash-dotted line) and TL pulses (black solid line). Right: Schematic view of the intensity profiles of pump (blue dashed line) and Stokes (red solid line) pulses used in the calculations of RAP (top), STIRAP (middle), and piecewise STIRAP (bottom). }
\vskip -.1truein
  \label{Fig-IntensityScan-AP}
\end{figure}

Our calculations do not indicate that the adiabatic techniques of two-photon RAP \cite{Malinovsky2001,book-Abragam}, STIRAP \cite{Scully2002} or piecewise STIRAP \cite{Shapiro2007a,Shapiro2008} are feasible. Indeed, AP is usually used to improve the robustness of an already high transfer rate. The left panel of Fig.~\ref{Fig-IntensityScan-AP} investigates the strong-field transfer for the three adiabatic schemes. The intensity profiles are shown in the right panel with wavelengths and intensities similar to those in Fig.~\ref{FigZ-IntensityScan}. For RAP, we show the transfer efficiency with pump and Stokes fields chirped in the opposite directions with $|\alpha'|=46,000$ fs$^2$. This shaping results in a frequency chirp of the two-photon pump-Stokes field. For STIRAP, transform-limited Stokes pulses precede transform-limited pump pulses by 130 fs. For piecewise STIRAP, we consider two pulse trains obtained via sinusoidal spectral phase modulation as described above and shifted relative to each other in the counter-intuitive order by the pulse train period \cite{Shapiro2007a,Shapiro2008}. In all three cases, the two-photon spectrum is lower than that of a TL pulse. Hence, the transfer efficiency at small intensities is lowered by the applied pulse shaping. At high intensities, the transfer efficiency is only slightly higher than that of transform-limited pulses, but still is significantly lower than that achieved by the non-adiabatic techniques in Fig.~\ref{FigZ-IntensityScan}.

Finally, we compare the results of the off-resonance excitation regime, with the two resonant cases, where a single-photon electronic excitation from either $v_0$ (case 1, $\l_p=573$ nm, $\l_S = 800$ nm) or $v_f$ (case 2, $\l_p=645$ nm, $\l_S = 800$ nm) is energetically allowed. The choice of these wavelengths is based on our experimental settings and on the previous studies \cite{Graefe2007,Graefe2007b}. In both cases, one-photon coupling to the excited electronic state at the pump frequency significantly complicates the population dynamics. The dependance of the transfer efficiency on the wavelength is highly irregular due to the interplay of time-dependent AC Stark shifts and a single-photon escape of population into  the $B$ manifold. In the first case, the overall transfer efficiency at high intensities is an order of magnitude lower than that in the far off-resonance arrangement considered in the main set of calculations. In the second case (previously studied in Ref.~\cite{Graefe2007,Graefe2007b}), the efficiency of the transfer can reach up to a few percent, i.e. slightly higher than that in the far off-resonance arrangement; however, the dependance on the laser parameters remains erratic. Here, the efficiencies observed in our calculations are substantially lower than those reported in Refs. \cite{Graefe2007,Graefe2007b}. In our calculation, the target population transfer into $v_f$ is hindered by one-photon coupling  to the excited electronic state at the pump frequency.


\section{Experimental study}
\subsection{Experimental setup}
A Ti:sapphire-based laser system (SpitFire Pro, Spectra-Physics) produces 2 mJ 130 fs pulses at 800 nm and 1 kHz repetition rate. These pulses serve as pump, whereas one optical parametric amplifier (TOPAS, Light Conversion) generates Stokes pulses at a wavelength of 973 nm and another OPA (OPA-800C, Spectra Physics) generates probe pulses at 578 nm.
\begin{figure}
\centering
 \includegraphics[width=1\columnwidth]{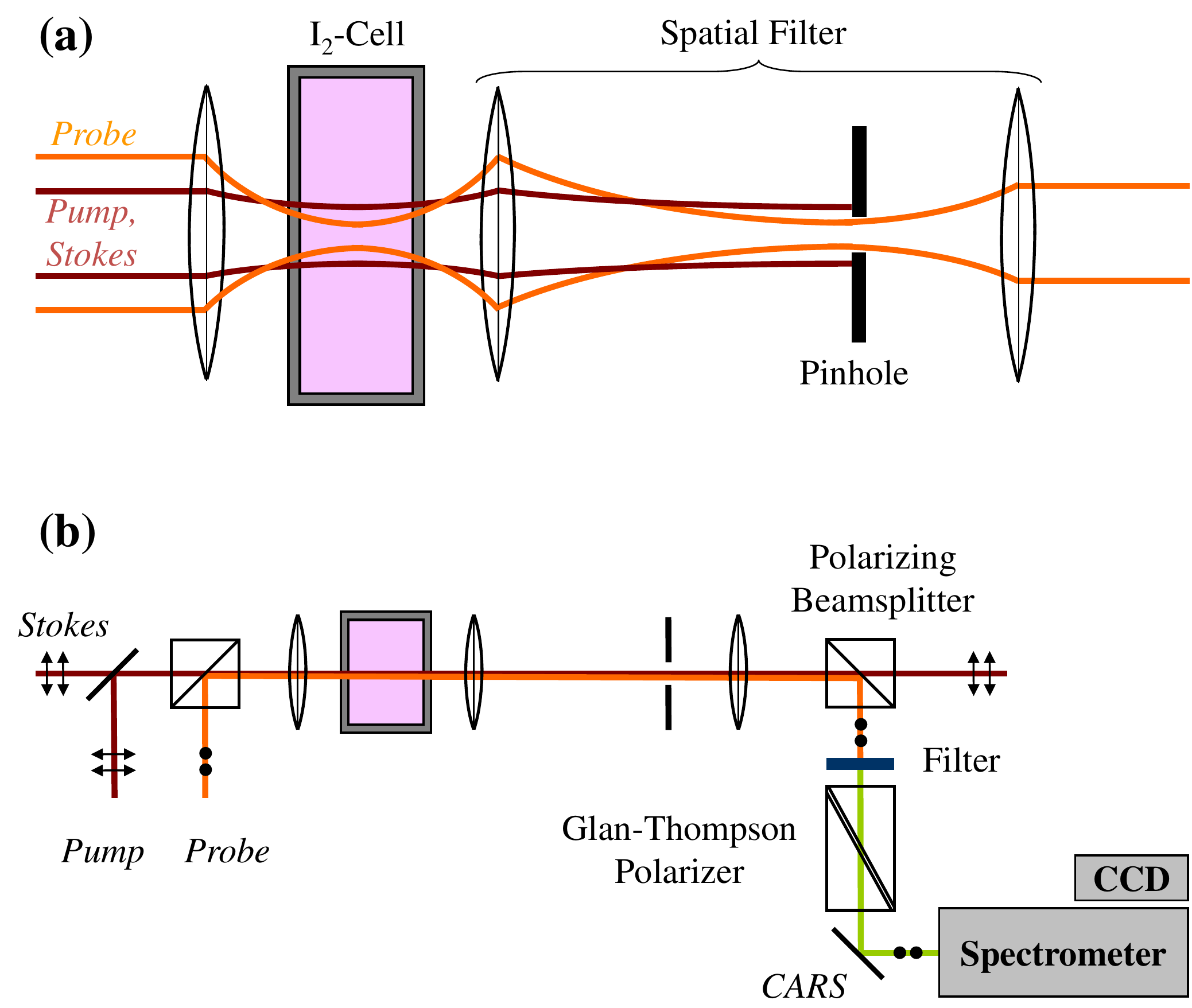}
     \caption{(Color online) Experimental setup: (a) Colinear CARS geometry with spatial filtering. (b) Use of polarizing optics and spectral filtering to separate CARS signal from the excitation pulses. }
  \vskip -.1truein
  \label{Fig:Setup}
\end{figure}

All three pulses are aligned collinearly and are spatially
overlapped inside a vapor cell filled with  I$_{2}$ at variable
temperature, see Fig.~\ref{Fig:Setup}(a). At 100$^\circ$C, used in
this work, the vapor pressure is 43 Torr \cite{Book-Yaws}.
Collinear geometry is implemented to increase the interaction
length and reduce the effect of spatial averaging over the
Gaussian beam profiles. The latter is also achieved by making the
focal size of the probe beam smaller than the size of the other
beams. With a 30 cm focusing lens, peak intensities of up to $5
\times 10^{13}$ W/cm$^2$ for pump and Stokes pulses and $2 \times
10^{12}$ W/cm$^2$ for probe pulses are obtained. Spatial filtering
in the detection channel is used to eliminate a majority of the
fluorescence background. In order to filter the output anti-Stokes
beam from the incident beams, we use polarization and spectral
filtering, as shown in Fig.~\ref{Fig:Setup}(b). Probe pulses are
linearly polarized in the orthogonal direction to pump and Stokes
pulses. Hence, the anti-Stokes polarization does not coincide with
that of the strong excitation fields, enabling one to block the
latter while passing through part of the CARS signal
\cite{Oron2003}. The filtered signal is sent to a spectrometer
equipped with a cooled ($-40^{\circ}$C) charge-coupled device
(CCD) camera. Two delay lines are used to vary the relative
arrival time of pump, Stokes and probe pulses. Pump and Stokes
pulses are shaped by two separate pulse shapers implemented in a
standard $4f$-geometry \cite{Weiner2000}.

\subsection{Results}
\begin{figure}
\centering
 \includegraphics[width=1\columnwidth]{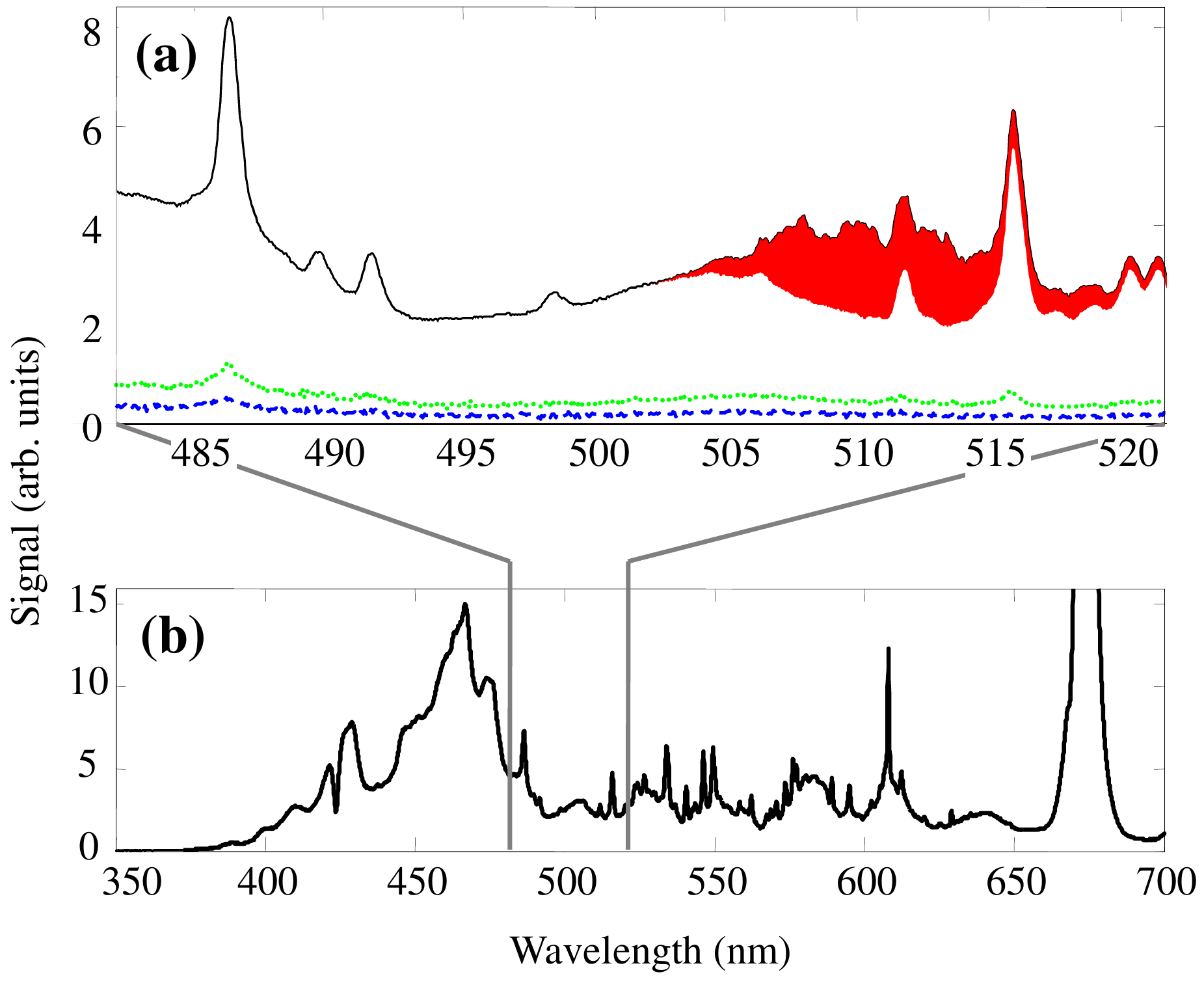}
     \caption{(Color online) Fluorescence spectrum of iodine due to strong pump and Stokes excitation. (a) Detected signal in the wavelength range from 482 nm to 522 nm stemming from TPA of Stokes (blue dashed), TPA of pump (green dotted) and TPA of mixed pump and Stokes (black solid). CARS signal is shown on top of the fluorescence curve (red shaded area). (b) Fluorescence spectrum in the range from 350nm to 700nm.}
  \vskip -.1truein
  \label{Fig:Fluorescence}
\end{figure}

To evaluate the effect of strong excitation fields, we start by detecting the output spectrum in the absence of probe pulses. The observed spectrum, shown in Fig.~\ref{Fig:Fluorescence}(b), is very broad and covers wavelengths above 400 nm. It corresponds to spontaneous emission following the process of two-photon absorption (TPA) in which any combination of two photons from the high intensity pump and Stokes fields are absorbed. The 400 nm cut-off is determined by the maximum energy from the absorption of two 800 nm pump photons. If only pump or Stokes photons are present, the fluorescence signal shows a quadratic dependency on the intensity of the corresponding laser beam. The fluorescence signal stemming from TPA of one pump and one Stokes photon shows a linear dependence on the intensity of both laser beams. It provides the main contribution to TPA (Fig.~\ref{Fig:Fluorescence}(a)). This is expected since the two-photon excitation amplitude scales as $(E_p+E_S)^2 = E_p^2 + E_S^2+2 E_pE_S$, resulting in a four-fold enhancement of the intensity dependence on $I_{p}I_{S}$ (Here $E_{p}$($I_{p}$) and $E_{S}$ ($I_{S}$) is the electric field amplitude (intensity) of pump and Stokes pulses, respectively). The exact proportionality factor is a function of the frequency dependent Frank-Condon factors. The strong peak at 678 nm is due to the non-resonant four-wave mixing process with two pump photons and one Stokes photon at frequency $2\wp-\ws$.

\begin{figure}
\centering
 \includegraphics[width=1\columnwidth]{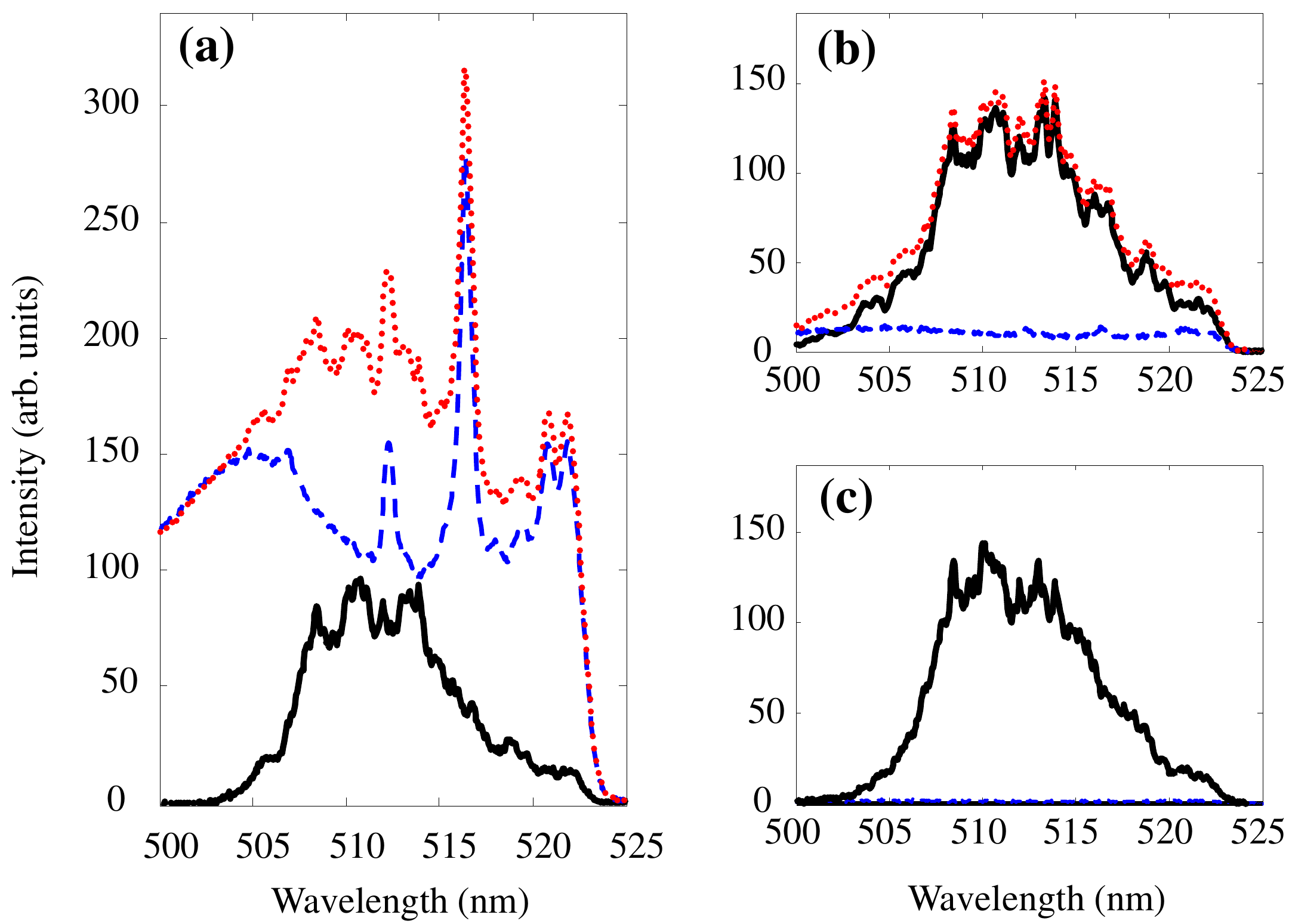}
     \caption{(Color online) Detected spectrum with strong pump and Stokes fields and a weak probe field (red dotted line), output spectrum in the absence of probe pulses (blue dashed line), CARS signal derived as the difference of the two spectra (black solid line); Pump and Stokes pulses are both (a) transform-limited, (b) modulated with a sinusoidal phase, (c) linearly chirped.}
  \vskip -.1truein
  \label{Fig:Fluorescence-Shaper}
\end{figure}

Strong two-photon fluorescence, comparable in magnitude to the detected CARS signal even after both spatial and polarization filters have been applied (Fig.~\ref{Fig:Fluorescence}(a)), reflects a high degree of two-photon coupling responsible for the detrimental Stark shifts. Suppressing the TPA-induced fluorescence with pulse shaping can be considered as an indirect evidence of lowering the influence of strong-field effects on the target Raman excitation. Both, the linear frequency chirp and the sinusoidal phase modulation, drastically reduce the fluorescence background due to the narrowing of the two-photon spectrum discussed earlier in the text. The effect is shown in Fig.~\ref{Fig:Fluorescence-Shaper}. As expected, linear chirping achieves better suppression of fluorescence, since the two-photon spectrum has fewer spectral components. In contrast to the suppressed TPA, the intensity of the CARS signal is rising, confirming the above argument about the mechanism of the strong-field induced saturation.

\begin{figure}
\centering
 \includegraphics[width=1\columnwidth]{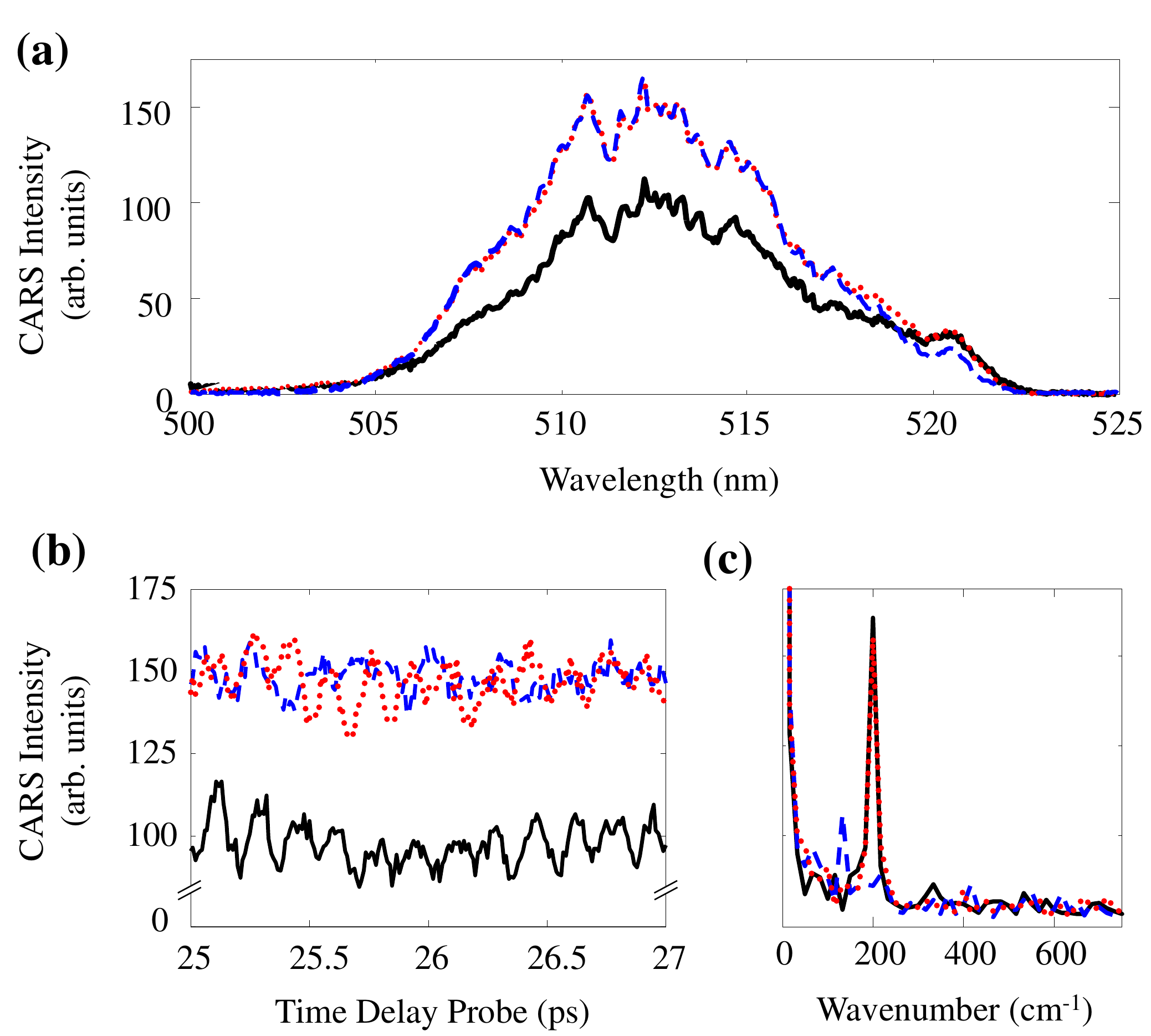}
     \caption{(Color online) Comparison of CARS intensity with different excitation schemes: Pump and Stokes pulses are both transform-limited (black solid line), linearly chirped (blue dashed line) or shaped with a sinusoidal phase mask (red dotted line). (a) CARS spectrum, (b) CARS signal as a function of the probe time delay and (c) the corresponding Fourier spectra.} \vskip -.1truein
  \label{Fig:ProbeScanSpectrum}
\end{figure}

In Fig.~\ref{Fig:ProbeScanSpectrum}, we plot the detected CARS signal obtained by subtracting the fluorescence background from the measured spectrum. We note that in this spectral region (i.e. around $\omega _{p} - \omega _{S} + \omega _{pr}$), the signal is free of non-resonant background since probe pulses are delayed by about 25 ps with respect to the temporal overlapping pump and Stokes pulses. The maximum available energy of the excitation pulses corresponds to peak intensities of $I_{p,S}=5 \times 10^{13}$ W/cm$^2$ of the unshaped TL pulses.  As can be seen in plots (a) and (b) of Fig.~\ref{Fig:ProbeScanSpectrum}, both phase masks (quadratic and sinusoidal) result in an almost identical signal enhancement by about 50\% in comparison to the case of the unshaped excitation.

The dependence of the CARS signal on the probe arrival time is plotted in Fig.~\ref{Fig:ProbeScanSpectrum}(b). Aside form the similar difference in the absolute signal strength, we note that the oscillations of the excited vibrational wave packet, clearly seen in the case of the TL excitation, are preserved in the case of the sinusoidal phase modulation and suppressed in the case of the linear frequency chirping. The Fourier spectra in Fig.~\ref{Fig:ProbeScanSpectrum}(c) reveal a peak at 200 cm$^{-1}$. This corresponds to a vibrational wave packet consisting of $v_{f}=10,11$ and 12 with an oscillation period of 167 fs. The peak is suppressed when frequency chirped pulses are used.
The effect is different from a well known feature of the weak-field regime, where the excitation amplitude is linearly proportional to the amplitude of the two-photon field at resonant frequencies. Although the sinusoidal phase modulation does not change the spectral amplitude at the frequencies of the available Raman transitions (see Fig.~\ref{Fig:RamanSpectrum}(b)), it results in a substantial signal gain while allowing us to excite a coherent wave packet. On the other hand, the observed signal enhancement with the frequency chirped pump and Stokes pulses is accompanied by the two-photon spectrum narrowing (Fig.~\ref{Fig:RamanSpectrum}(a)), and a consequent loss of the wave packet oscillations. %

\begin{figure}
\centering
 \includegraphics[width=0.8\columnwidth]{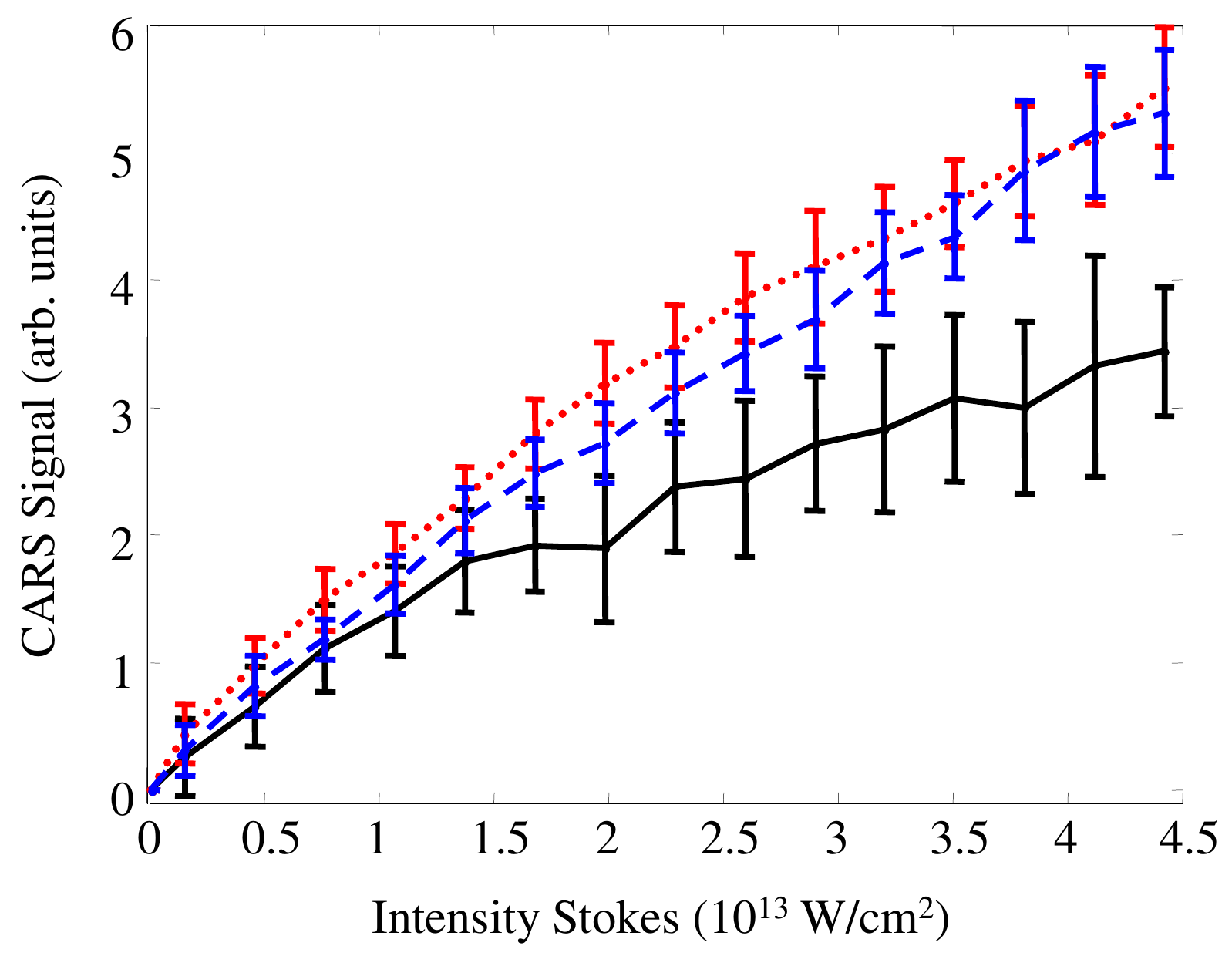}
     \caption{(Color online)
     CARS intensity as a function of Stokes intensity with $I_p=5 \times  10^{13}$W/cm$^2$.
     Line assignments are the same as in Fig.~\ref{Fig:ProbeScanSpectrum}.}
  \vskip -.1truein
  \label{Fig:EnergyScan}
\end{figure}

By scanning the intensity of the Stokes beam in
Fig.~\ref{Fig:EnergyScan}, we observe the effect of pulse shaping
above approximately $1.5 \times  10^{13}$W/cm$^2$ when the signal
from the unshaped transform-limited excitation starts to slow
down. We attribute this onset of saturation to the deleterious
strong-field effects, which are suppressed when the pulses are
shaped. In the latter case, the signal keeps growing almost
linearly exceeding the unshaped limit by about 50\%. Our numerical
analysis suggests that frequency chirping should provide the
highest CARS signal (see Fig.~\ref{FigZ-IntensityScan}), whereas
in the experiment, the two spectral masks result in a similar
performance. In the simulations, we set the Stokes wavelength to
match the peak of the two-photon spectrum with a Raman resonance.
In the experiment, however, the wavelengths cannot be changed
easily and the frequencies are matched by introducing a delay
between pump and Stokes pulses (see also next paragraph). This
delay reduces the two-photon spectral amplitude, since it is
shifted from the maximum of the two-photon spectrum, thus lowering
the signal. In the case of the sinusoidal phase modulation, the
exact overlap of the two-photon spectrum with the Raman transition
frequencies is easily maintained by adjusting the central
modulation frequencies $\tilde{\w}_{p,S}$ of
Eq.~\ref{SinusoidalPhase} with pump and Stokes pulse shapers.

Following our numerical analysis, we examine the techniques of RAP
and STIRAP in both the off-resonance and near-resonance cases. In
the off-resonant case, i.e. with the same choice of pump and
Stokes wavelengths as before ( $\lambda_p=800$ nm, $\lambda_S=973$
nm), we find no evidence of the AP enhancement of CARS signal. As
required for rapid adiabatic passage, we chirp pump and Stokes
fields in opposite directions with $|\alpha'|=50,000$ fs$^2$, and
observe CARS intensity always decreasing below the TL limit. This
agrees with the perturbative, rather than adiabatic, description
of the Raman process. As discussed earlier in the text,
opposite-sign chirping of pump and Stokes pulses results in a
lower two-photon intensity which is not sufficient for initiating
AP.

\begin{figure}
\centering
 \includegraphics[width=0.9\columnwidth]{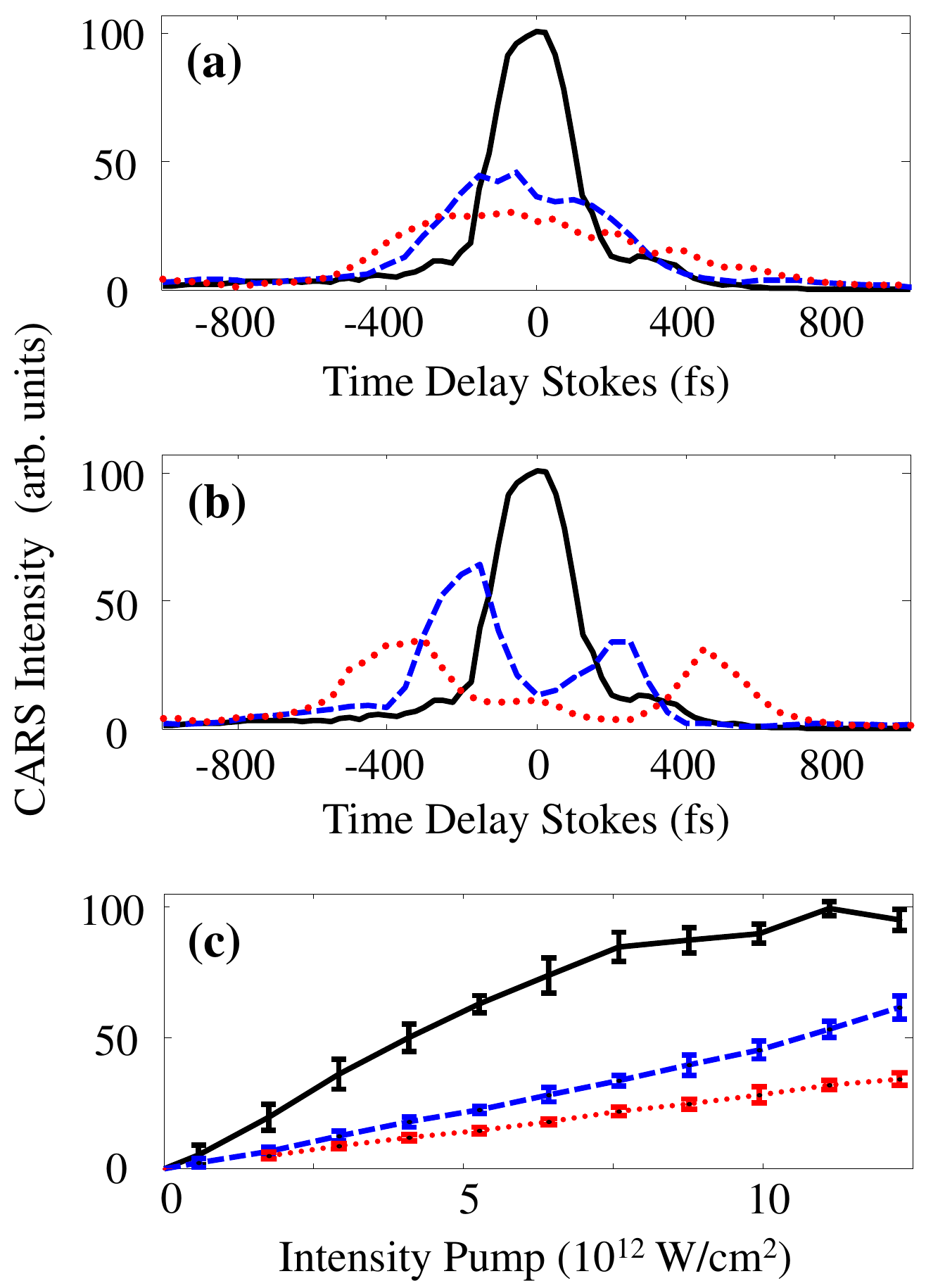}
\caption{(Color online) Near-resonance CARS. (a) CARS intensity as a function of pump-Stokes time delay for TL pulses (black solid line) and oppositely chirped pulses with $\alpha'_{S,p}=\pm10,000$ fs$^2$ (blue dashed line) and $\alpha'_{S,p}=\pm25,000$ fs$^2$ (red dotted line). (b) CARS intensity as a function of pump-Stokes time delay and (c) as a function of pump intensity with $I_S=4 \times  10^{12}$W/cm$^2$ for TL pulses (black solid line) and same-sign linearly chirped pulses with $\alpha'_{S,p}=-10,000$~fs$^2$ (blue dashed line) and $\alpha'_{S,p}=-25,000$~fs$^2$ (red dotted line).}
\vskip -.1truein
  \label{Fig:AP-close-to-resonance}
\end{figure}

In order to test the non-perturbative regime necessary for AP, we shift the central excitation wavelengths to $\l_p=645$ nm and $\l_S = 800$ nm. This choice of wavelengths replicates that of Refs.~\cite{Graefe2007,Graefe2007b}. In Fig.~\ref{Fig:AP-close-to-resonance}(a), the spectra of the excitation pulses are chirped linearly in opposite directions (RAP) with two different chirp amplitudes $\alpha'$. CARS intensity is plotted as a function of the time delay between the pulses. We find no signatures of RAP (i.e. signal increase with frequency chirping at zero time delay) or STIRAP (i.e. signal increase with time delay at zero frequency chirp), and always record the strongest CARS signal  with  transform-limited pump and Stokes pulses perfectly overlapping in time. From this observation, we conclude that no adiabatic evolution takes place.

In Fig.~\ref{Fig:AP-close-to-resonance}(b,c), we apply the successful same-sign frequency chirping, found in the off-resonance case, to the near-resonance interaction scheme. As discussed earlier, the two-photon spectrum becomes narrower and, depending on the exact pump-Stokes time delay, only one Raman transition is covered at a time (c.f. Fig.~\ref{Fig:RamanSpectrum}(a)). This is reflected by the peaks in the CARS intensity at certain time delays seen in Fig.~\ref{Fig:AP-close-to-resonance}(b). Fig.~\ref{Fig:AP-close-to-resonance}(c) shows that for all beam intensities achievable in the experiment, the overall CARS intensity is lower than that for the unshaped pulses. Unlike the off-resonance case, pulse shaping does not improve the overall CARS efficiency. We attribute the difference to complicated dynamics due to one-photon coupling to the excited electronic state.

\section{Summary and discussion}
In this experimental and theoretical work, we investigated the
utility of femtosecond pulse shaping to increase the efficiency of
CARS in the strong-field excitation regime. We showed that when
molecular eigenstates are significantly modified by the applied
laser fields, the efficiency of exciting molecular vibration is no
longer achieved with unshaped transform-limited pulses. We
analyzed two qualitatively different approaches to enhancing the
magnitude of non-resonant vibrational excitation with shaped
pulses.

In the first approach, pulse shaping was used to suppress
strong-field effects, such as AC Stark shifts, while preserving
the amplitude of the two-photon field at the frequency of Raman
resonances. Numerically, we showed that linear frequency chirping
of strong pump and Stokes pulses can provide several times better
excitation efficiency than that achieved with transform-limited
pulses. The efficiency of Raman excitation with sinusoidal phase
modulation showed an intermediate result. In the experiment, both
techniques improved the strength of the observed CARS signal by
50\% with respect to the TL case. Sinusoidal phase modulation
resulted in a substantial signal gain while allowing us to excite
a coherent vibrational wave packet, which may prove important in
strong-field spectroscopic applications.

Multiple sets of pump and Stokes wavelengths were tested. Our main
study was devoted to the far off-resonance case, where the
dependence of vibrational excitation on the field parameters
allows reasonably simple interpretation. In the near-resonance
case, when either the ground or the target state were coupled to
the excited electronic state, the population dynamics were
erratic. In spite of the higher two-photon Raman matrix elements,
the excitation efficiency in all studied near-resonance cases was
either negligibly higher or even substantially lower (depending on
the choice of wavelengths) than in the far off-resonance case.

In the second approach, we put to test three adiabatic passage scenarios which excel in a robust and efficient population transfer in simple two- or three-level atomic systems -- RAP, STIRAP and piecewise STIRAP \cite{Vitanov2001,Shapiro2007a,Shapiro2008}. All AP schemes studied in this work failed to establish an adiabatic evolution and did not improve the efficiency of Raman excitation. We associate this lack of success with two factors. Our numerical analysis showed that the complexity of the molecular spectrum prevents reaching the AP regime which is typically manifested by a high amount of population transferred to the target state or set of states. From the experimental point of view, adiabatic passage requires shaping the excitation pulses in a way which reduces the two-photon spectral power density, making the available laser power insufficient for establishing an adiabatic evolution.

\begin{acknowledgements}
The authors would like to thank Vladimir Malinovsky and Sergey Zhdanovich for valuable discussion. This work has been supported by the DTRA grant HDTRA1-09-1-0021.

\end{acknowledgements}



\end{document}